\documentstyle[epsfig]{aipproc}
\begin{document}
\title{EGRET (GeV) Blazars}

\author{R. Mukherjee}
\address{$^*$ Dept. of Physics \& Astronomy, Barnard College, Columbia
University,\\
 New York, NY 10027}

\maketitle

\begin{abstract}

The EGRET instrument on the Compton Gamma Ray Observatory (1991-2000) has 
positively detected high energy $\gamma$-ray emission from more than 67 
active galaxies of the blazar class. 
The majority of the EGRET blazars are flat-spectrum radio quasars, which 
are characterized by inferred isotropic luminosities often as high as 
$3\times 10^{49}$ ergs s$^{-1}$. The remainder are BL Lac objects, some of 
which have been detected at TeV energies ($> 250$ GeV) by  
ground-based atmospheric Cherenkov telescopes.  
One of the remarkable characteristics observed 
in blazars is that the $\gamma$-ray luminosity often dominates the 
bolometric power in these sources. 
The detection of blazars by EGRET has undoubtedly been one of the highlights 
of the mission, and has forever impacted our understanding of the emission 
mechanisms in these objects. In this article, we summarize the properties of 
EGRET blazars, and review the constraints that the EGRET observations place 
on the various models of $\gamma$-ray production in these sources. 

\end{abstract}

\section*{Introduction}

The earliest surveys of the sky above 100 MeV were carried out by 
the COS-B satellite, which detected $\gamma$-rays from 3C 273 
\cite{swan78}, the first extragalactic $\gamma$-ray source. 
At the time EGRET was launched in 1991, 3C 273 
was the only extragalactic source known to emit $\gamma$-rays. The subject of 
$\gamma$-ray blazars got its first dramatic boost after the EGRET detection of  
several high energy $\gamma$-ray blazars \cite{mont95,mukh97,hart99}. 
EGRET discovered two remarkable 
characteristics of these sources: (1) In the majority of the blazars, the 
$\gamma$-ray emission dominates the bolometric power in the spectral energy 
distributions of these objects, and (2) blazars are characterized by 
short-time scale variabilities, often on the order of days to months. 

Almost all the active galaxies detected by EGRET belong to the blazar-class 
of objects. 
Although the term ``blazar'' is not well-defined, it includes OVV quasars, 
BL Lac objects, and high polarization quasars (HPQs). 
These sources are 
characterized by emissions that include high radio and optical polarizations, 
strong variability at all wavelengths, 
and non-thermal, continuum spectra. Blazars detected by EGRET 
are radio-loud, flat-spectrum radio sources, 
with radio spectral indices $\alpha_r\geq -0.6$ 
for a flux density of $S_r \propto \nu_r^{\alpha_r}$. 
Several blazars have been observed to 
exhibit apparent superluminal motion (3C 279, 3C 273, 3C 454.3, PKS 0528+134,
for example \cite{verm94}), as evidenced from VLBI radio observations. 
EGRET observations of blazars are summarized in section I. 

Three of the EGRET blazars (Mrk 421, Mrk 501, PKS 2155-304) have now been
detected at TeV energies ($> 250$ GeV) by ground-based atmospheric Cherenkov 
telescopes (ACTs) \cite{cata99}. These results have reinvigorated the 
$\gamma$-ray studies of blazars and have opened a new chapter in high 
energy astrophysics. As new ACTs with lower energy thresholds ($< 50$ GeV ) 
become operational \cite{ongr98}, 
the high energy $\gamma$-ray studies of blazars will be
conducted from the ground before the next generation satellite-based 
experiments like GLAST \cite{gehr99} or AGILE \cite{tava99} 
come on line. The future of high energy 
$\gamma$-ray blazar studies looks very promising. 
Some of the EGRET blazars observed at TeV energies are 
described in section II. 

EGRET observations of blazars have helped constrain models of $\gamma$-ray 
emission mechanisms in these objects. 
The combination of high luminosity, short-term variability, and 
superluminal motion observed in blazars all bolster arguments for 
relativistic beaming in these objects. This is in 
agreement with the strong association of EGRET blazars with radio-loud 
flat-spectrum radio sources, with many of them showing superluminal 
motion in their jets. Most blazar models adopt a beaming
paradigm. EGRET observations of blazars have enabled us to test several
competing blazar emission models, most of which can be divided into two broad
classes: ``leptonic'' and ``hadronic.'' It is generally believed that the 
broadband spectral energy distribution of blazars is due to energy losses of 
highly energetic particles via both synchrotron and Compton processes. 
A summary of the current blazar models and their implications, as pertaining to
EGRET sources, is given in section III. The possible contributions of 
blazars to the diffuse extragalactic $\gamma$-ray background are discussed in 
section IV. 

This article presents a snap shot of the current knowledge on EGRET-detected 
$\gamma$-ray blazars and selectively reviews some of the interesting highlights of
the field. A comprehensive review is not possible, owing to the limited number
of the allotted pages. Instead, the reader is referred to several recent 
articles which serve as excellent introductions to the exciting field of 
$\gamma$-ray blazars detected by EGRET \cite{mont95,mukh97,hart97,mukh98}. 

\section{EGRET Observations}

This article reviews the blazars listed in the third EGRET (3EG) catalog 
\cite{hart99}. 
These blazars were typically observed by EGRET for a period of about 2
weeks. Some observations were as short as 1 week, while others were as long 
as 3 to 5 weeks. EGRET's threshold sensitivity ($> 100$ MeV) for a 
single 2-week observation was 
$\sim 3\times 10^{-7}$ photons cm$^{-2}$ s$^{-1}$. Details of the EGRET
instrument, and data analysis techniques are given elsewhere
\cite{thom93,hart99}.  

Of the 271 sources detected by EGRET above 100 MeV, 67 are blazars detected 
with a high degree of confidence (a detection of at least a $4\sigma$ 
statistical significance at high galactic latitudes, and a $5\sigma$ 
detection for $\vert b\vert <10^\circ$ \cite{hart99}). In addition, 27 
other blazars are classified in the 3EG catalog as possible AGN \cite{hart99}. 
This is because their identification is questionable, either because the 
candidate 
source has a low radio flux, or lies outside the 95\% error contour of the 
EGRET source. This makes 
blazars the second-largest class of EGRET sources, following the 
``unidentified'' sources that have no known counterparts at other wavelengths. 
EGRET has detected blazars with redshifts ranging from $\sim$ 0.03 to 2.28. 

The majority of the blazars detected by EGRET are classified as flat-spectrum 
radio quasars (FSRQs). The 46 FSRQs with strong identifications, listed in the 
3EG catalog, are shown in 
Table 1. Of these, 0440-003, CTA 102, and 3C 454.3 are not strictly
``flat-spectrum'' in the range between 1.4 and 5 GHz, or 2.7 and 8 GHz. 
The BL Lac objects detected by EGRET are listed in Table 2. Not included in 
the table is Mrk 501, which is discussed in section II. The BL Lac objects are
characterized by stronger polarization and weaker optical lines than FSRQs,
which makes redshift determination more difficult. Other differences between
FSRQs and BL Lac objects are discussed in the following sections. 
Not listed in either table are 
four flat-spectrum radio sources 0616-116, 0738+545,
1759-396, and 1908-201, also detected by EGRET. In addition, EGRET has 
detected $\gamma$-ray emission from Cen A, a radio galaxy, the only one seen to
emit $\gamma$-rays \cite{sree99a}. 

\begin{table}
\scriptsize
\caption{Flat-Spectrum Radio Quasars (FSRQs) detected by EGRET}
\label{table:1}
\begin{tabular}{@{}cccccccc}
\tableline
Name  &Other Name &       $l$   &    $b$  &Flux$\times 10^{-8}$ & Photon            & $z$    & $L_{48}$ \\ 
          &           &             &         &ph cm$^{-2}$s$^{-1}$ & Spectral Index    &        & ergs s$^{-1}$ \\ 
\tableline

0202+149 &  4C+15.05  &      147.95 & -44.32  &$  23.6  \pm   5.6$  &$  2.23\pm   0.28$ &  0.405 & 0.0600 \\
0208-512 &  PKS       &      276.10 & -61.89  &$  85.5  \pm   4.5$  &$  1.99\pm   0.05$ &  1.003 & 1.4500 \\
0336-019 &  CTA026    &      188.40 & -42.47  &$ 118.8  \pm  22.0$  &$  1.84\pm   0.25$ &  0.852 & 1.4000 \\
0414-189 &            &      213.90 & -43.29  &$  49.5  \pm  16.1$  &$  3.25\pm   0.68$ &  1.536 & 2.1600 \\
0420-014 &            &      194.88 & -33.12  &$  50.2  \pm  10.4$  &$  2.44\pm   0.19$ &  0.915 & 0.6900 \\
0430+285 &            &      170.48 & -12.58  &$  22.0  \pm   2.8$  &$  1.90\pm   0.10$ &    -   &   -    \\
0440-003 &  NRAO190   &      197.39 & -28.68  &$  79.0  \pm  10.1$  &$  2.37\pm   0.18$ &  0.844 & 0.9100 \\
0446+112 &            &      187.86 & -20.62  &$ 109.5  \pm  19.4$  &$  2.27\pm   0.16$ &  1.207 & 2.7900 \\
0454-463 &            &      252.40 & -38.40  &$   7.7  \pm   2.1$  &$  2.75\pm   0.35$ &  0.858 & 0.0900 \\
0459+060 &            &      193.99 & -21.66  &$  12.1  \pm   3.1$  &$  2.36\pm   0.40$ &  1.106 & 0.2500 \\
0458-020 &            &      201.35 & -25.47  &$  11.2  \pm   2.3$  &$  2.45\pm   0.27$ &  2.286 & 1.1900 \\
0528+134 &            &      191.50 & -11.09  &$  93.5  \pm   3.6$  &$  2.46\pm   0.04$ &  2.060 & 7.8600 \\
0827+243 &            &      199.91 &  31.69  &$  24.9  \pm   3.9$  &$  2.42\pm   0.21$ &  2.046 & 2.0600 \\
0836+710 &            &      143.49 &  34.79  &$  10.2  \pm   1.8$  &$  2.62\pm   0.16$ &  2.172 & 0.9700 \\
0954+556 &            &      159.55 &  47.33  &$   9.1  \pm   1.6$  &$  2.12\pm   0.18$ &  0.901 & 0.1200 \\
1156+295 &  4C+29.45  &      201.53 &  78.63  &$  50.9  \pm  11.9$  &$  1.98\pm   0.22$ &  0.729 & 0.4300 \\
1222+216 &            &      254.91 &  81.53  &$  13.9  \pm   1.8$  &$  2.28\pm   0.13$ &  0.435 & 0.0400 \\
1226+023 &  3C273     &      289.84 &  64.47  &$  15.4  \pm   1.8$  &$  2.58\pm   0.09$ &  0.158 & 0.0100 \\
1229-021 &            &      292.58 &  59.66  &$  12.7  \pm   2.9$  &$  2.85\pm   0.30$ &  1.045 & 0.2400 \\
1243-072 &            &      300.96 &  55.99  &$   9.8  \pm   2.1$  &$  2.73\pm   0.17$ &  1.286 & 0.2900 \\
1253-055 &  3C279     &      304.98 &  57.03  &$ 179.7  \pm   6.7$  &$  1.96\pm   0.04$ &  0.538 & 0.7800 \\
1331+170 &            &      346.29 &  76.68  &$   9.4  \pm   2.7$  &$  2.41\pm   0.47$ &  2.084 & 0.8100 \\
1406-076 &            &      334.23 &  50.30  &$  97.6  \pm   9.1$  &$  2.29\pm   0.11$ &  1.494 & 4.0000 \\
1424-418 &            &      321.66 &  16.98  &$  29.5  \pm   5.3$  &$  2.13\pm   0.21$ &  1.522 & 1.2600 \\
1510-089 &            &      351.49 &  40.37  &$  18.0  \pm   3.8$  &$  2.47\pm   0.21$ &  0.361 & 0.0300 \\
1606+106 &  4C+10.45  &       23.51 &  41.05  &$  34.9  \pm   5.6$  &$  2.63\pm   0.24$ &  1.226 & 0.9200 \\
1611+343 &            &       55.44 &  46.29  &$  26.5  \pm   4.0$  &$  2.42\pm   0.15$ &  1.401 & 0.9400 \\
1622-297 &            &      348.67 &  13.38  &$ 258.9  \pm  15.3$  &$  2.07\pm   0.07$ &  0.815 & 2.7700 \\
1622-253 &            &      352.28 &  16.37  &$  42.6  \pm   6.6$  &$  2.21\pm   0.13$ &  0.786 & 0.4200 \\
1633+382 &  4C+38.41  &       61.21 &  42.26  &$ 107.5  \pm   9.6$  &$  2.15\pm   0.09$ &  1.814 & 6.8000 \\
1725+044 &            &       27.27 &  20.62  &$  17.9  \pm   4.1$  &$  2.67\pm   0.26$ &  0.296 & 0.0200 \\
1730-130 &  NRAO530   &       12.00 &  10.57  &$  36.1  \pm   3.4$  &$  2.23\pm   0.10$ &  0.902 & 0.4800 \\
1739+522 &            &       79.37 &  32.05  &$  18.2  \pm   3.5$  &$  2.42\pm   0.23$ &  1.375 & 0.6200 \\
1741-038 &            &       22.19 &  13.42  &$  21.9  \pm   5.3$  &$  2.42\pm   0.42$ &  1.054 & 0.4100 \\
1830-210 &            &       11.92 &  -5.50  &$  26.6  \pm   3.7$  &$  2.59\pm   0.13$ &  1.000 & 0.4500 \\
1933-400 &            &      358.65 & -25.23  &$  21.9  \pm   4.9$  &$  2.86\pm   0.40$ &  0.966 & 0.3400 \\
1936-155 &            &       23.95 & -17.12  &$  55.0  \pm  18.6$  &$  3.45\pm   1.27$ &  1.657 & 2.8400 \\
2022-077 &            &       36.72 & -24.40  &$  74.5  \pm  13.4$  &$  2.38\pm   0.17$ &  1.388 & 2.5900 \\
2052-474 &            &      352.56 & -40.20  &$  23.6  \pm   6.0$  &$  2.04\pm   0.35$ &  1.489 & 0.9600 \\
2209+236 &            &       81.83 & -25.65  &$  14.6  \pm   4.2$  &$  2.48\pm   0.50$ &     -  &  -     \\
2230+114 &  CTA102    &       77.45 & -38.50  &$  19.2  \pm   2.8$  &$  2.45\pm   0.14$ &  1.037 & 0.3500 \\
2251+158 &  3C454.3   &       86.05 & -38.30  &$  53.7  \pm   4.0$  &$  2.21\pm   0.06$ &  0.859 & 0.6500 \\ 
2320-035 &            &       76.82 & -58.07  &$  38.2  \pm  10.1$  &   --              &  1.411 & 1.3800 \\ 
2351+456 &            &      113.39 & -15.82  &$  14.3  \pm   3.7$  &$  2.38\pm   0.38$ &  1.992 & 1.1200 \\
2356+196 &            &      107.01 & -40.58  &$  16.0  \pm   4.7$  &$  2.09\pm   0.35$ &  1.066 & 0.3100 \\
\tableline
\end{tabular}
\end{table}

\begin{table}
\scriptsize
\caption{BL Lac Objects detected by EGRET}
\label{table:2}
\begin{tabular}{@{}cccccccc}
\tableline
Name  &Other Name &       $l$   &    $b$  &Flux$\times 10^{-8}$ & Photon            & $z$    & $L_{48}$ \\ 
          &           &             &         &ph cm$^{-2}$s$^{-1}$ & Spectral Index    &        & ergs s$^{-1}$ \\ 
\tableline

0219+428 &  3C66A     &      140.22 & -16.89  &$ 18.7  \pm  2.9$  &$  2.01\pm   0.14$ &   0.444 & 0.0500 \\
0235+164 &  OD+160    &      156.46 & -39.28  &$ 65.1  \pm  8.8$  &$  1.85\pm   0.12$ &   0.940 & 0.9500 \\
0454-234 &            &      223.96 & -34.98  &$ 14.7  \pm  4.2$  &$  3.14\pm   0.47$ &   1.009 & 0.2500 \\
0537-441 &            &      250.08 & -30.86  &$ 25.3  \pm  3.1$  &$  2.41\pm   0.12$ &   0.894 & 0.3300 \\
0716+714 &            &      143.98 &  28.00  &$ 17.8  \pm  2.0$  &$  2.19\pm   0.11$ &   0.300 & 0.0200 \\
0735+178 &            &      202.16 &  17.88  &$ 16.4  \pm  3.3$  &$  2.60\pm   0.28$ &   0.424 & 0.0400 \\
0829+046 &            &      219.60 &  23.82  &$ 16.8  \pm  5.1$  &$  2.47\pm   0.40$ &   0.180 & 0.0100 \\
0851+202 &  OJ+287    &      207.19 &  35.43  &$ 10.6  \pm  3.0$  &$  2.03\pm   0.35$ &   0.306 & 0.0100 \\
0954+658 &            &      145.78 &  43.11  &$ 15.4  \pm  3.0$  &$  2.08\pm   0.24$ &   0.368 & 0.0300 \\
1101+384 &  Mrk421    &      179.97 &  65.04  &$ 13.9  \pm  1.8$  &$  1.57\pm   0.15$ &   0.031 & 0.0002 \\
1219+285 &  ON+231    &      197.27 &  83.52  &$ 11.5  \pm  1.8$  &$  1.73\pm   0.18$ &   0.102 & 0.0015 \\
1334-127 &            &      320.07 &  46.95  &$ 11.8  \pm  3.4$  &$  2.62\pm   0.42$ &   0.539 & 0.0500 \\
1604+159 &  4C+15.54  &       29.18 &  43.84  &$ 42.0  \pm 12.3$  &$  2.06\pm   0.41$ &   0.357 & 0.0800 \\
2032+107 &            &       56.12 & -17.18  &$ 13.3  \pm  3.1$  &$  2.83\pm   0.26$ &   0.601 & 0.0700 \\
2155-304 &            &       17.45 & -52.23  &$ 30.4  \pm  7.7$  &$  2.35\pm   0.26$ &   0.116 & 0.0100 \\
2200+420 &  BLLac     &       92.56 & -10.39  &$ 39.9  \pm 11.6$  &$  2.60\pm   0.28$ &   0.069 & 0.0024 \\
\tableline
\end{tabular}
\tablenotetext{Note that the table lists sources only from the 3EG catalog, and does not
include  Mrk 501, which was not detected by EGRET until later. The EGRET
detection of this source is described in section II. }
\end{table}

\subsection{Time Variability of EGRET Blazars}

\begin{figure}
\centerline{\psfig{file=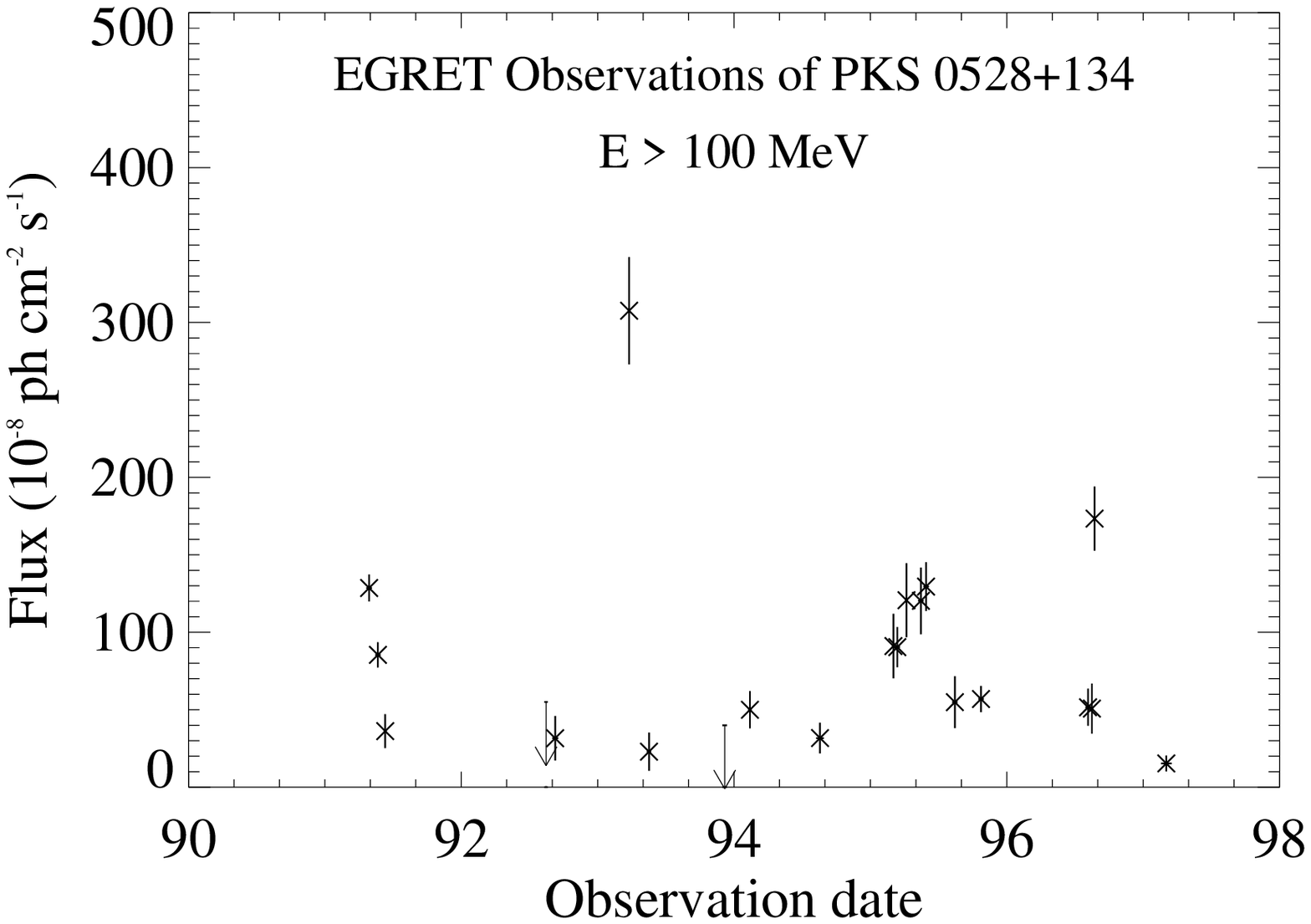,height=2.5in,width=3.5in,bbllx=0pt,bblly=360pt,bburx=555pt,bbury=710pt,clip=.}\psfig{file=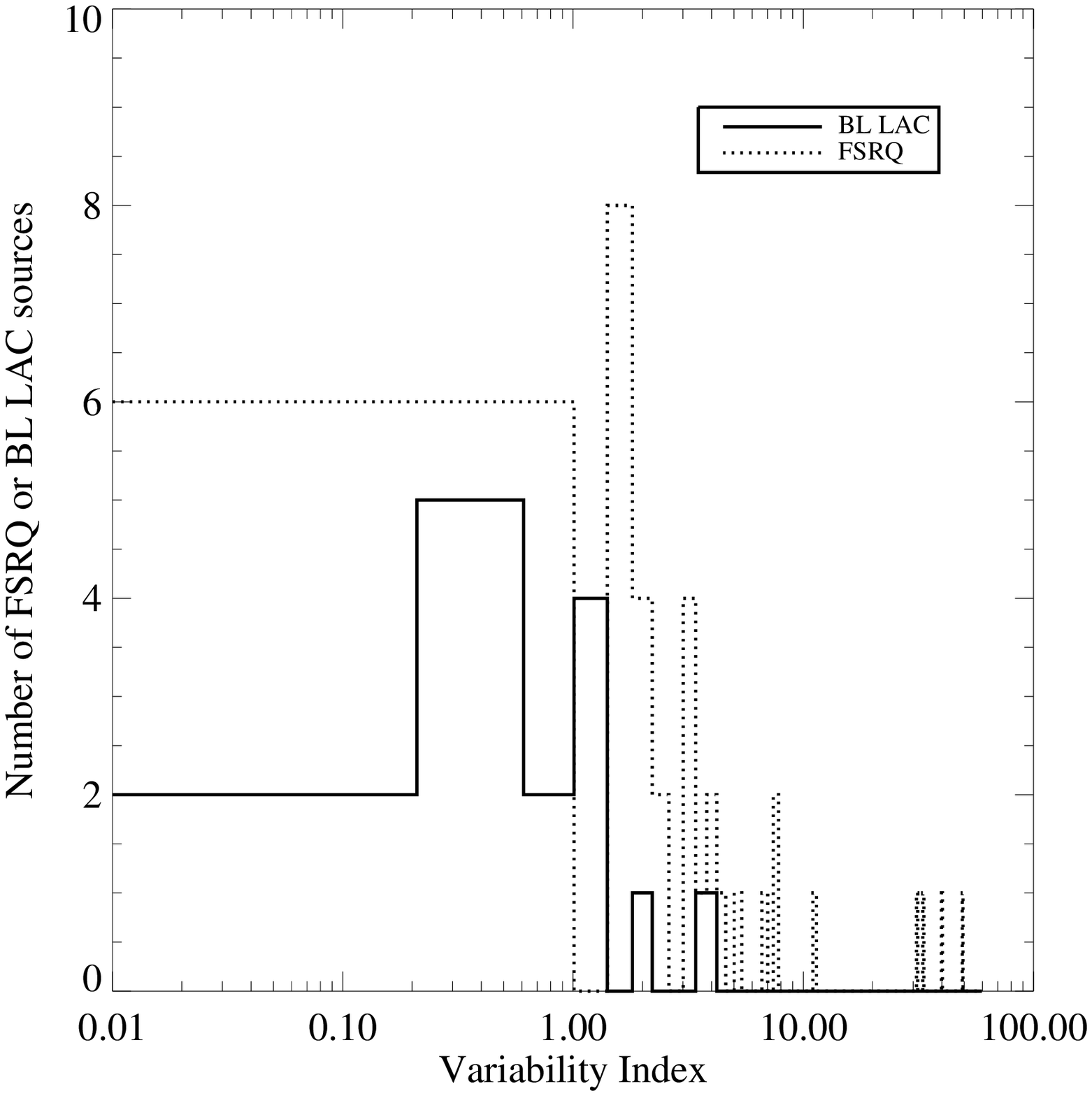,height=2.5in,width=3.0in,bbllx=65pt,bblly=100pt,bburx=575pt,bbury=615pt,clip=.}}
\caption{(Left) Flux history of the blazar PKS 0528+134. (Right) Distributions
of variability indices for FSRQs (dashed line) and BL Lac objects 
(solid line). }
\label{fluxhist}
\end{figure}

Blazars detected by EGRET have exhibited time variability in their $\gamma$-ray 
emission on timescales of months, to sometimes well under a day. Long term 
flux 
variability for EGRET blazars have been discussed earlier by several authors 
\cite{mont95,mukh97}. Figure~\ref{fluxhist} shows the flux history of the blazar PKS
0528+134 over a 
period of six years \cite{mukh99}. 
Each data point corresponds to the time averaged flux from
the source during a viewing period (VP). 
Upper limits, at the 95\% confidence level, are shown for any 
detections below the $2\sigma$ level. The figure demonstrates variability on the
time scale of months. PKS 0528+134 is one of the most variable sources detected
by EGRET. A dramatic flare was seen from the source in 1993 March when its flux
was several times greater than that of the Crab \cite{mukh99}. 

The flux variability of the blazars can be quantified by assigning a 
``variability index,'' $V =\log Q$, where $Q=1-P_{\chi}(\chi^2,\nu)$
\cite{mcla96}. The quantity $P_{\chi}$ is the probability of observing 
$\chi^2$ or something larger from a $\chi^2$ distribution with $\nu$ degrees of
freedom. The $\chi^2$ values were obtained by fitting the data similar to that
shown in Figure~\ref{fluxhist} to a
constant flux, using a least square fit method. The quantity $V$ has been used
in the past to assess the strength of variability \cite{mcla96,mukh97}, following the somewhat
arbitrary classification: non-variable - $V< 0.5$; 
variable - $V\ge 1$; uncertain - $0.5\le V<1$. 
Figure~\ref{fluxhist} shows a distribution of the variability indices of the FSRQs and BL
Lac objects listed in tables 1 and 2. Although the sample of the blazars is
small, the figure indicates that FSRQs are generally more variable than the BL
Lac objects, in the EGRET energy range. Nearly 75\% of all FSRQs may be
classified as variable. In comparison, most of the unidentified EGRET sources
are found to be non-variable, although recent studies have found short time
scale variability in a few unidentified EGRET sources \cite{wall00}. 

Although the study of short time scale variability is limited for most blazars 
by the small number of photons detected by EGRET during a single observation
period, some EGRET blazars have been observed to flare dramatically on the time
scale of days. For example, PKS 1622-297 exhibited a major flare during the
observation in 1995. A flux increase by a factor of at least 3.6 was observed 
in a 
period of less than 7 hours \cite{matt97}. The peak flux observed was $(17\pm
3)\times 10^{-6}$ cm$^{-2}$ s$^{-1}$ ($E > 100$ MeV), which
corresponded to an isotropic luminosity of $2.9 \times 10^{49}$ ergs
s$^{-1}$\cite{matt97}. The characteristic time scale of variability, or the
``doubling time,'' can be
somewhat arbitrarily defined as the shortest timescale during which the source
flux is seen to change by a factors of 2. In the case of PKS 1622-297, the
doubling time was less than 3.8 hr. A dramatic flare was also observed 
in the blazar 3C 279, in which the doubling time was $\sim 8$ hours 
\cite{wehr97}. Other blazars that have shown strong variability on timescales 
of 1 to 3 days in the past are PKS 0528+134, 3C 454.3, PKS 1633+382, 1406-076,
and CTA26. Studies of fast variations of $\gamma$-ray emission in blazars 
using structure function analysis techniques have also been carried out for a 
few selected sources \cite{nand97,wagn97}. 

The study of short time scale variability in blazars is important as they help
to constrain the size of emission regions in these objects. A flux variation by
a factor of two on an observed time scale $\delta t_{obs}$ limits the size $r$
of the emitting region to roughly $r\le c\delta t_{obs}/(1+z)$ using simple 
light travel time arguments. This implies that the blazar emission region is 
very compact. 

\subsection{Spectra of EGRET blazars} 

\begin{figure}[b!]
\centerline{\psfig{file=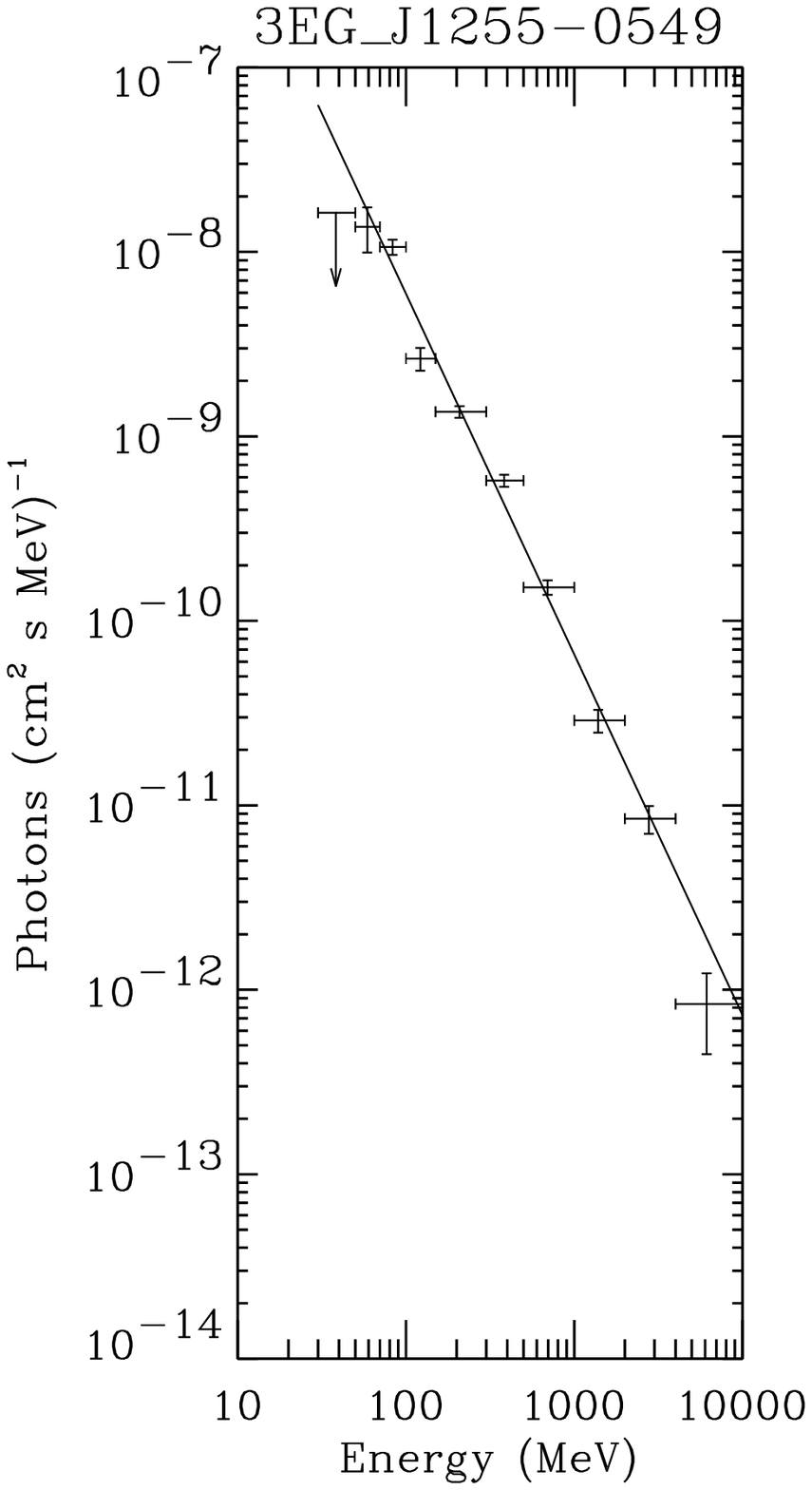,height=4.0in,bbllx=160pt,bblly=190pt,bburx=450pt,bbury=720pt,clip=.}\psfig{file=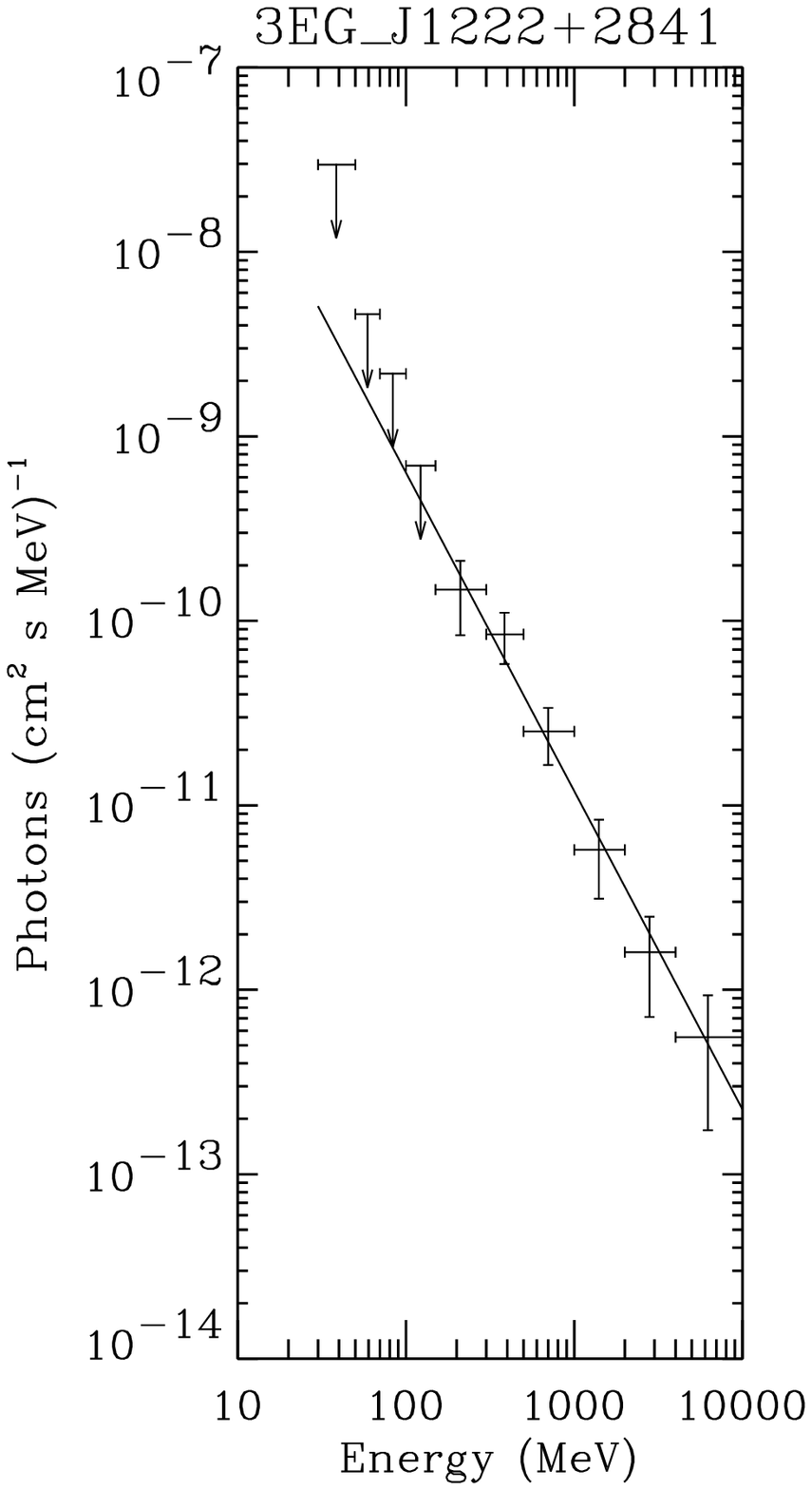,height=4.0in,bbllx=160pt,bblly=190pt,bburx=450pt,bbury=720pt,clip=.}}
\caption{Photon spectra of the FSRQ 3C 279 (left) and the BL Lac object 
1219+285 (right) in the energy range 30 MeV to 10 GeV. The solid lines are
the best fit to a power law. $2\sigma$ upper limits are shown as downward
arrows.}
\label{egretspec}
\end{figure}

The spectra of blazars detected by EGRET cover the energy range from 30 MeV to
10 GeV and are well-described by a simple power-law model of the form 
$F(E)=k(E/E_0)^{-\alpha}$ photons cm$^{-2}$ s$^{-1}$ MeV$^{-1}$, 
where the photon spectral index, $\alpha$, and the coefficient, $k$, 
are the free parameters. The energy normalization factor, $E_0$, is 
chosen so that the statistical errors in the power law index and the overall 
normalization are uncorrelated. Details of the EGRET spectral analysis 
techniques for blazars may be found elsewhere \cite{mont95}. 
Figure~\ref{egretspec} shows the spectra of two blazars detected by EGRET 
\cite{hart99}. 
The spectral indices of all the blazars detected by EGRET are included in 
tables 1 and 2. The average photon spectral index in the EGRET 
energy band, assuming a simple power law fit to the spectrum, is $\sim 2.2$. 
There is no evidence of a spectral cutoff for energies below 10 GeV.  
Figure~\ref{spec} shows a plot of spectral indices versus redshift 
of all blazars detected by EGRET. 

\subsection{Luminosity of EGRET blazars}

\begin{figure}[b!]
\centerline{\psfig{file=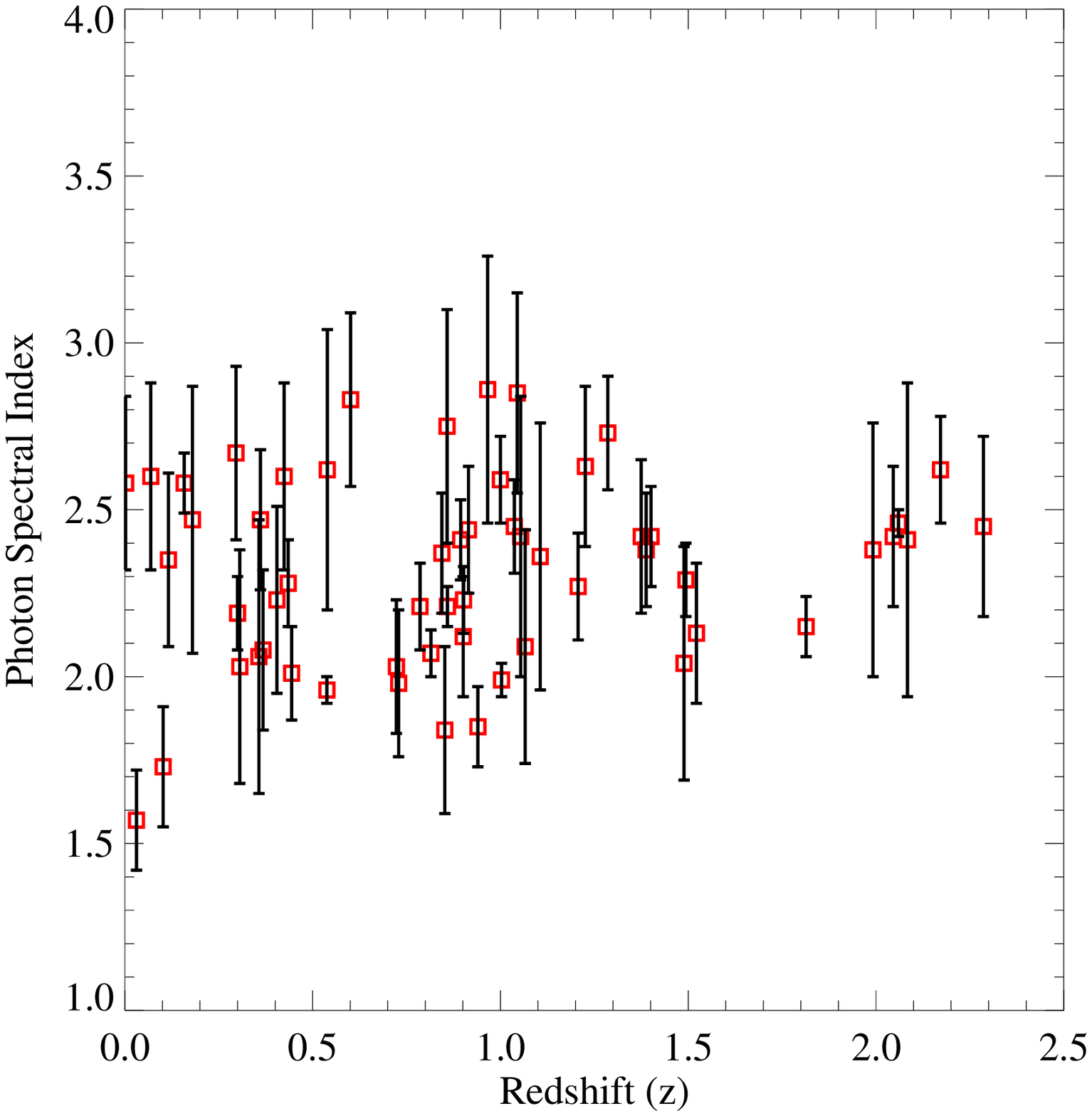,height=3.0in,bbllx=50pt,bblly=150pt,bburx=555pt,bbury=620pt,clip=.}}
\centerline{\psfig{file=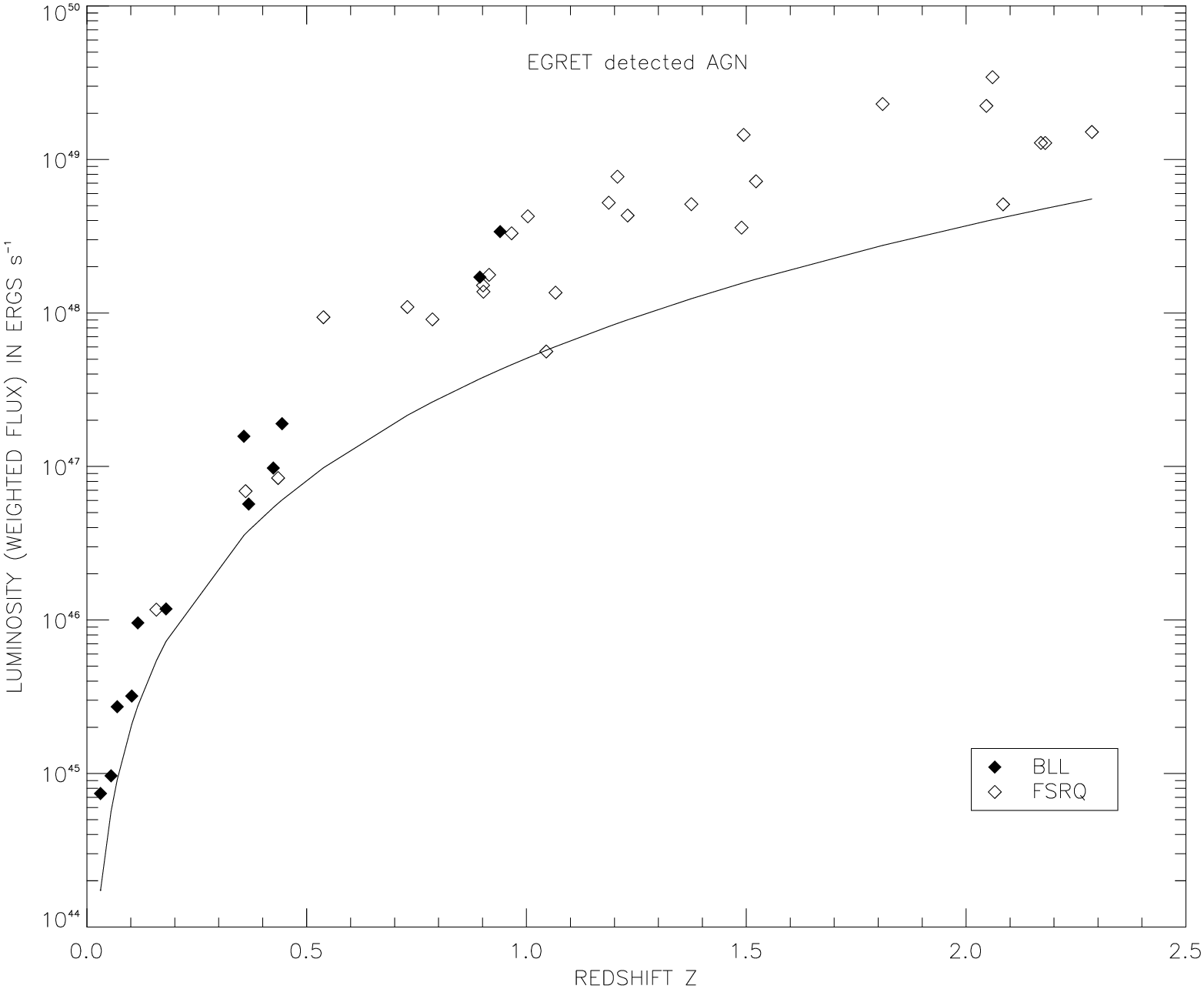,height=3.5in,bbllx=70pt,bblly=700pt,bburx=650pt,bbury=1300pt,clip=.}}
\vspace{10pt}
\caption{(Top) Spectral index versus redshift for EGRET blazars. (Bottom) Luminosity vs redshift for 
blazars detected by EGRET. The BL Lac objects are indicated with filled
symbols. The typical detection threshold for EGRET is shown as a solid curve.}
\label{spec}
\end{figure}

Tables 1 and 2 list the $\gamma$-ray luminosity of the blazars, calculated assuming
isotropic emission. The $\gamma$-ray luminosity was estimated using the method
described in \cite{mukh97}. Figure~\ref{spec} shows a plot of the luminosity as a
function of redshift  for the FSRQs and BL Lac objects detected by EGRET. The 
solid line corresponds to the typical detection threshold for EGRET as a
function of redshift, for relatively good conditions. As seen from the figure,
the BL Lac objects (indicated with dark diamonds) are predominantly closer and
have lower luminosities than the FSRQs. 

\section{SEDs of EGRET blazars}

\subsection{FSRQs and Radio-selected BL Lac Objects}

In order to decipher the underlying physics of the emission mechanisms in
blazars, it is essential to study the broad band spectral energy distributions
(SEDs) of these sources. Since EGRET first started detecting blazars, several
multiwavelength campaigns have been organized to simultaneously observe these
sources at different wavelengths (see \cite{shra97} for a recent review). 
Although these campaigns have proven difficult to organize, and the SEDs of
blazars are often non-simultaneous or under-sampled, they have helped constrain
blazars models. 

\begin{figure}
\centerline{\psfig{file=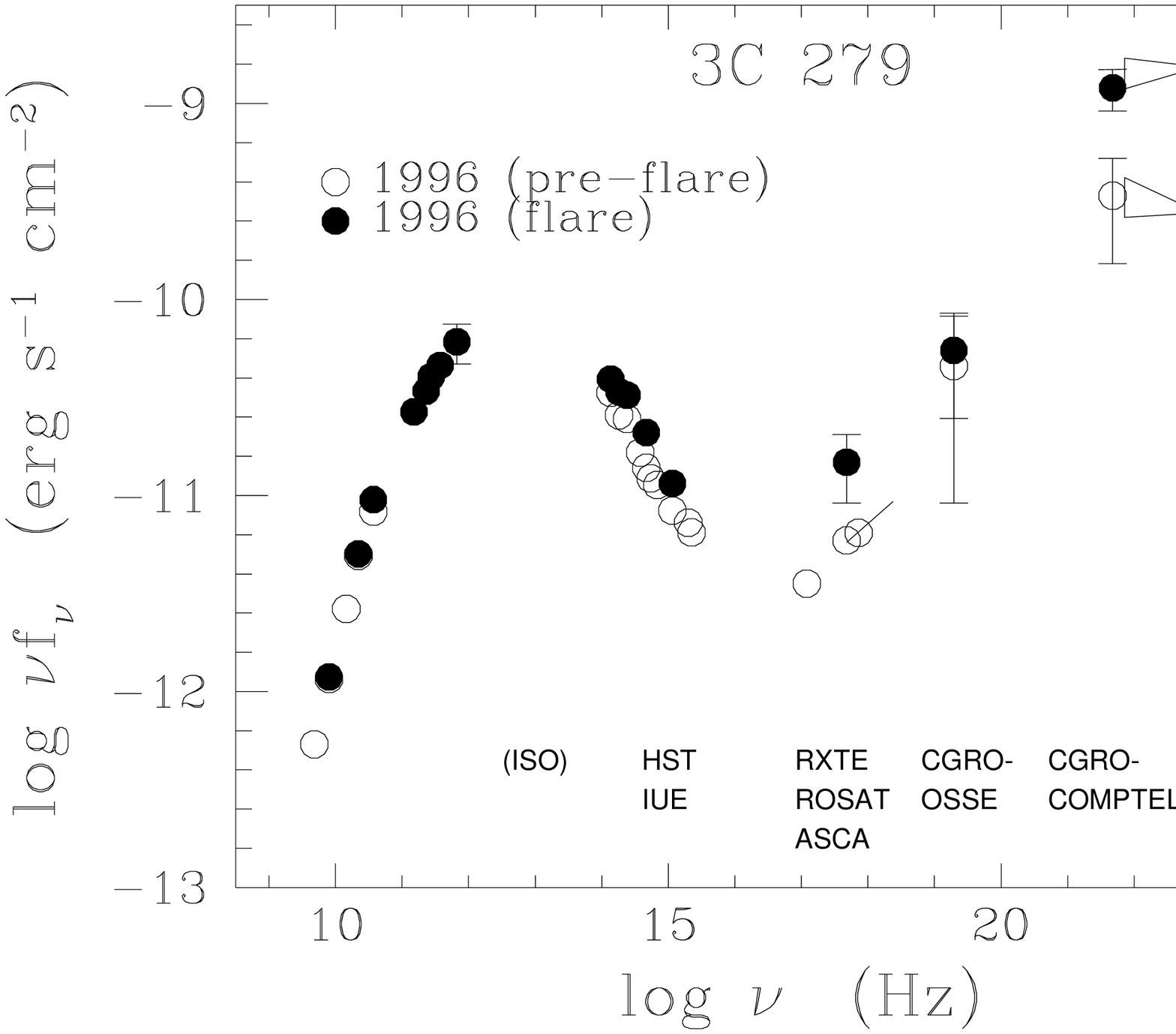,height=2.5in,bbllx=35pt,bblly=25pt,bburx=725pt,bbury=540pt,clip=.}\psfig{file=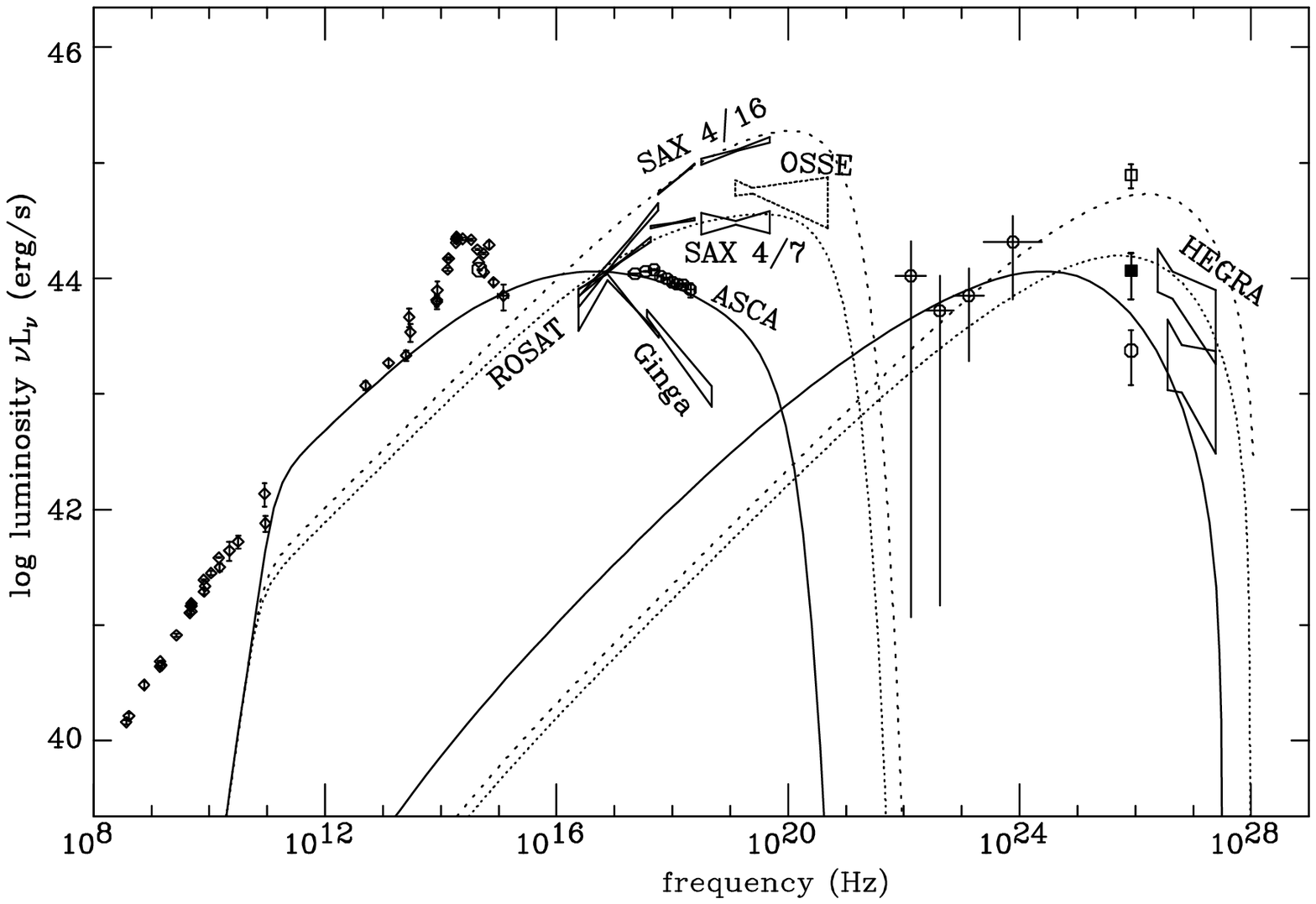,height=2.5in,width=3.0in,bbllx=35pt,bblly=215pt,bburx=540pt,bbury=560pt,clip=.}}
\caption{(Left) SED of 3C 279 from radio to $\gamma$-ray energies for different
epochs. The figure demonstrates the considerable variability that is often seen
in the broad band spectra of blazars. (Right) Broadband SED of MrK 501 during several different epochs. The filled 
and open squares correspond to Whipple flux on April 7 and 16, 1997,
respectively. The circles correspond to data taken in 1996 March. The diamonds 
correspond to non-simultaneous data from the NED data base. (See text for
references.)  }
\label{sed}
\end{figure}

The SED of the blazar 3C 279 from radio to $\gamma$-ray energies over 15 decades 
in energy for two different epochs \cite{wehr97} is shown in 
Figure~\ref{sed}. 
3C 279 is one of the most intensely monitored of all EGRET blazars. The
figure plots the quantity $\nu F_{\nu}$ as a function of frequency, thus showing
the relative amounts of power detected in each energy band. The overall 
SEDs of both FSRQs and BL Lac objects show two broad components: the 
synchrotron component and the inverse Compton (IC) 
component. In the case of FSRQs,
as shown in Figure~\ref{sed}, 
the first peak is in the optical-IR band, while the second peak is
in the MeV-GeV energy range. The figure
demonstrates the dominance of the $\gamma$-ray luminosity over that at other
wavebands. 3C 279 also shows considerable spectral variability, particularly in
the $\gamma$-ray band, between different epochs. This has been found to be a
characteristic feature of the FSRQs observed by EGRET. 

SEDs of the FSRQs and BL Lac objects detected by EGRET have been studied by 
many researchers, both for individual sources as well as for source classes. 
Some recent work may be found in \cite{samb99,foss99,ghis98,urry99}. 

\subsection{Blazars detected at TeV Energies}

Of the blazars detected by EGRET, only three have been seen at TeV energies
($> 250$ GeV), all three being X-ray selected BL Lac (XBL)  objects. These are 
Mrk 421, Mrk 501, and PKS 2155-304 \cite{week99}. Figure~\ref{sed} 
shows the broad band SED of 
Mrk 501 measured during the outburst in 1997 May \cite{kata99}. Mrk 501 was 
detected first at TeV energies before it was recognized in the EGRET data. 
The strongest EGRET detection of the source occurred in 1996 
May, when Mrk 501 was detected at 5.2~$\sigma$ at energies above 500 MeV 
\cite{sree99b}. Six high energy photons ($> 1$ GeV) were observed from the
direction of Mrk 501 
in a 1-day interval, when the probability for this happening by chance  
was $< 10^{-6}$ \cite{sree99b}.  A power-law fit to the EGRET data suggested a 
spectral index of $1.2\pm 1.0$ for Mrk 501, 
and is the hardest known blazar spectrum at 
GeV energies \cite{sree99b}.  

The SED of Mrk 501 shown in Figure~\ref{sed} is similar to that of other X-ray selected
BL Lac objects. Unlike in FSRQs, the $\gamma$-ray peak in the SED of this
high-frequency-peaked BL Lac (HBL) no longer
dominates the spectrum, although it plays a very important part. The 
synchrotron peak of Mrk 501 is located in the X-ray band, while the Compton
peak  is in the GeV-TeV regime. 
In fact, a correlation between the peak
frequencies of the synchrotron and IC components of the SED and 
the total energy density of the emitting region has been noted
previously \cite{ghis98}. There appears to be evidence for a 
well-defined sequence in the properties of different blazar classes, 
namely, HBL, LBL (low frequency-peaked BL Lacs), 
HPQ (high-polarization quasars), and LPQ 
(low-polarization quasars). The trends in the observed properties 
are a decrease in frequencies of the synchrotron and inverse Compton 
peaks, and an increase in the power-ratio of the high and low energy 
spectral components along the sequence HBL $\to$ LBL $\to$ HPQ $\to$ LPQ.

At a redshift of 0.031, Mrk 421 is the closest XBL seen by EGRET, and 
the first one to be detected by ground-based ACTs \cite{cata99}. Although 
it is a rather weak source at GeV energies, it has been persistently detected 
by EGRET \cite{liny92}. Above 30 MeV the source has a very hard photon spectral
index of $1.57\pm 0.15$, and has a high energy peak in its SED at TeV 
energies. 

\section{Gamma ray emission models}

\subsection{Leptonic models}
It is generally believed 
that blazars are powered by accretion of matter on to a supermassive black
hole, and that $\gamma$-ray emission originates in strongly beamed jets. The
absence of intrinsic $\gamma-\gamma$ pair absorption in the observed 
blazar spectra strongly points to the fact that $\gamma$-ray emission is beamed
radiation from the jet. A review of the constraints placed on AGN models from
the $\gamma$-ray observations may be found in \cite{schl96}.  

Leptonic jet models explain the radio to UV continuum from blazars as
synchrotron radiation from high energy electrons in a relativistically 
outflowing jet which has been ejected from an accreting supermassive 
black hole \cite{blan79}. The emission in the MeV-GeV 
range is believed to be due to the inverse Compton scattering of 
low-energy photons by the same relativistic electrons in the jet. 
However, two main issues remain questionable: the source of the soft 
photons that are inverse Compton scattered, and the structure of the inner 
jet, which cannot be imaged directly. The soft photons can originate as 
synchrotron emission from within the jet (the {\sl synchrotron-self-Compton} 
or SSC process), or from a nearby accretion disk, 
or they can be disk radiation reprocessed in broad-emission-line
clouds (the {\sl external radiation Compton} or the ERC process). 
A review on leptonic jet models and the relevant references 
may be found in \cite{boet99a}. 

Recently, Mukherjee et al. (1999) have applied a leptonic jet model to the 
SED of PKS 0528+134 observed by EGRET from 1991 to 1997 \cite{mukh99}. A 
combination of SSC and ERC models was used. A
two-component model, in which the target photons are produced 
externally to the $\gamma$-ray emitting region, but also including 
an SSC component, was required to suitably reproduce the spectral 
energy distributions of the source. The analysis indicated that 
there is a trend in the observed properties of PKS 0528+134, as 
the source goes from a $\gamma$-ray low state to a flaring state. 
The model fits presented in \cite{mukh99} indicated that 
during the higher $\gamma$-ray states, the bulk Lorentz factor of 
the jet increased and the ERC component dominated the high-energy
emission. The energies of the electrons giving rise to the synchrotron
peak decreased, and the power-ratio of the $\gamma$-ray and low energy
spectral components increased, as the source went from a low to a high 
$\gamma$-ray state. 

\begin{figure}[t!]
\centerline{\epsfig{file=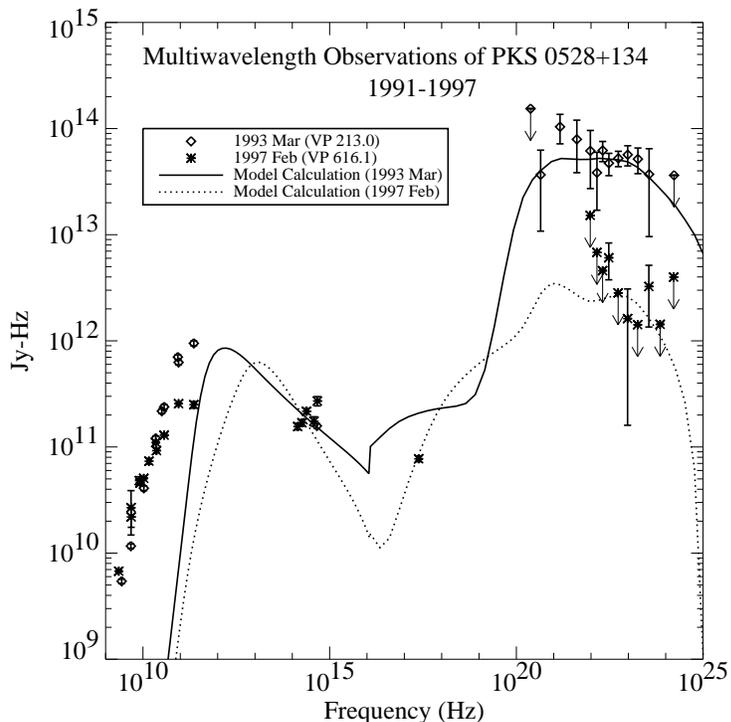,height=3.8in,bbllx=50pt,bblly=110pt,bburx=560pt,bbury=625pt,clip=.}}
\caption{ Data and model calculations for the broad band spectra of 
PKS 0528+134 during its low (1997 Feb) and flaring (1993 March) states. }
\label{0528}
\end{figure}

In Figure~\ref{0528} 
the fit results to two extreme states of PKS 0528+134 from \cite{mukh99} 
are compared. During the 1993 March observations, PKS 0528+134 was in its 
highest state ever, while the 1997 February data correspond to one of the 
lowest states of the source during the EGRET observations. 
The source exhibits considerable spectral
variability between different epochs, a characteristic typical of many FSRQs 
observed by EGRET. The model calculations of the SED of PKS 0528+134 yielded 
a shift of the synchrotron peak frequency towards 
lower frequencies in a $\gamma$-ray bright state of the source, as has been 
predicted analytically \cite{boet99}. The spectral 
variability in PKS 0528+134 appears to arise from the different 
Doppler boosting patterns of the SSC and the ERC radiations. The 
relative contributions of the SSC and ERC cooling mechanisms seem 
to be related to the optical to $\gamma$-ray flux ratio from the source. The 
SSC mechanism plays a larger role if the source is in a low flux state. The 
ERC mechanism is the dominant cooling mechanism when the source is in a 
high $\gamma$-ray state. 

B\"ottcher \cite{boet99} has demonstrated that the predicted peak shift 
is characteristic of a leptonic jet model in which the $\gamma$-ray
emission is dominated by the ERC process during $\gamma$-ray flares, 
but does not depend on the physical details of such a model. 
This is particularly important, as the SEDs of most blazars are still quite 
poorly constrained observationally, and can in many cases be 
fitted equally well with several fundamentally different models. 

A similar analysis was recently performed for the multi-epoch SEDs of 
3C 279 \cite{hart00}, where the same general trend in the SEDs were observed 
as in the case of PKS 0528+134. 

\subsection{Hadronic models}

In contrast to leptonic jet models where the jet consists of a hot
$e^\pm$ plasma, in hadronic models the observed $\gamma$-ray emission is
initiated by hot protons interacting with ambient gas or low-frequency 
radiation. In the {\sl proton induced cascade} (PIC) model \cite{mann93,mann96}
the relativistic protons interact electromagnetically and hadronically, i.e. 
through photo-pion production, with low frequency synchrotron radiation. Such
a model would require extremely high luminosity in high energy protons in 
order to explain the observed TeV $\gamma$-ray fluxes in blazars. 
In contrast, in the 
{\sl synchrotron proton blazar} models \cite{ahar00,muck00}, the high energy 
emission in the blazars detected at TeV energies is attributed to synchrotron 
radiation of extremely high energy protons. In such models, synchrotron 
radiation becomes the dominant energy loss mechanism of the high
energy protons. {\sl Synchrotron proton blazar} models have recently been
applied to the case of the EGRET blazars Mrk 501 and Mrk 421 
\cite{ahar00,prot00}. 

\section{Blazars and the extragalactic gamma-ray background}

The precise origin of the extragalactic diffuse emission is not well-known and 
possibly includes contributions from diffuse origins as well as unresolved 
point sources.  
Recently, the evolution and luminosity function of the EGRET blazars was 
used to estimate the contribution to the diffuse extragalactic background 
\cite{chia98}. It was found that the evolution was
consistent with a pure luminosity evolution. 
The redshift distribution of EGRET blazars was used to characterize the 
low end of the luminosity function better. 
The high end of the luminosity function was fixed by a non-parametric 
estimate \cite{chia98}. 
Using the lower limit of the de-evolved luminosity function, the
$\gamma$-ray loud AGN contribution to the extragalactic $\gamma$-ray flux
was estimated to be $4.0^{+1.0}_{-0.9}\times 10^{-6}$ photons cm$^{-2}$
s$^{-1}$ sr$^{-1}$. The sky-averaged flux contribution of identified EGRET
blazars is $\simeq 1\times 10^{-6}$ photons cm$^{-2}$ s$^{-1}$ sr$^{-1}$.
Contribution to the diffuse background by unresolved blazars, therefore, was
found to be $\sim 3.0^{+1.0}_{-0.9}\times 10^{-6}$ photons 
cm$^{-2}$ s$^{-1}$ sr$^{-1}$. The estimated extragalactic diffuse flux for 
$E>100$ MeV is $\simeq 1.36\times 10^{-5}$ photons cm$^{-2}$ s$^{-1}$
sr$^{-1}$ \cite{sree98}. These calculations indicate that 
blazars cannot account for all of the
diffuse extragalactic $\gamma$-ray background at the energies considered. 
Under this scenario, 
only $\sim$ 25\% of the diffuse extragalactic emission
measured by SAS-2 and EGRET can be attributed to unresolved $\gamma$-ray
blazars.  These findings are consistent with other studies \cite{muck98}, and 
contrary to estimates which assume a linear 
correlation between the measured radio and $\gamma$-ray fluxes and attribute a 
much larger contribution by blazars to the $\gamma$-ray background 
(e.g. \cite{stec96}). 
These results lead to the exciting conclusion that other sources of diffuse
extragalactic $\gamma$-ray emission must exist. 

\section{Summary and Future Prospects}

During its lifetime EGRET detected more than 67 blazars at energies above 100
MeV. This has been one of the most remarkable contributions from the mission. 
Only a 
handful of the EGRET sources have been detected with ground-based ACTs at
energies above 250 GeV, although most of them have been observed 
\cite{kerr95}. This 
indicates that spectral changes must exist in the unexplored range between 
50 and 250 GeV, possibly due to intrinsic absorption at the source or
due to $\gamma$-$\gamma$ pair production off the extragalactic background light
(EBL) \cite{goul66}. 
In the future, it will be important to complement blazar studies by 
space-based experiments such as AGILE or GLAST, with ground-based instruments 
sensitive to the energy range 50 to 250 GeV. Ground-based instruments like 
STACEE and CELESTE with energy thresholds lower than 250 GeV are currently
operational and well-suited to the study of several northern hemisphere AGN. 
Other experiments like VERITAS, HESS, MAGIC and CANGAROO III will soon come 
online. A review of these experiments, and a few others, and 
the relevant references may be found in \cite{ongr98,week99}. 

\bigskip
I wish to thank M. B\"ottcher, R. C. Hartman, P. Sreekumar, for providing some
of the data and figures presented here and at the Heidelberg 
$\gamma$ 2000 conference. This research was supported in part
by the National Science Foundation and the Research Corporation.


\begin{references}
\bibitem{swan78} Swanenburg, B. N., et al., Nature, 275 (1978) 298. 
\bibitem{mont95} C. von Montigny, et al., ApJ, 440 (1995) 525. 
\bibitem{mukh97} R. Mukherjee, et al., ApJ, 490 (1997) 116. 
\bibitem{hart99} R. C. Hartman, et al., ApJ, 123 (1999) 79.  
\bibitem{verm94} R. C. Vermeulen \& M. H. Cohen, ApJ, 430 (1994) 467. 
\bibitem{cata99} M. Catanese \& T. C. Weekes, PASP review, 
astro-ph 9906501 (1999). 
\bibitem{ongr98} R. A. Ong, Phys. Rep., 305 (1998) 95. 
\bibitem{gehr99} N. Gehrels \& P. Michelson, APh, 526 (1999) 297. 
\bibitem{tava99} M. Tavani, et al., A\&AS, 138 (1999) 569. 
\bibitem{hart97} R. C. Hartman, et al., AIP Conf. Proc. 410 (1997) 307. 
\bibitem{mukh98} R. Mukherjee, in ``Observational Evidence for Black Holes in
the Universe,'' ed. S. K. Chakrabarti (1999) 215, Kluwer Academic Publishers. 
\bibitem{thom93} D. J. Thompson, et al., ApJS, 102 (1995) 259. 
\bibitem{sree99a} P. Sreekumar, et al., APh, 11 (1999) 221. 
\bibitem{mukh99} R. Mukherjee, et al., ApJ, 527 (1999) 132. 
\bibitem{mcla96}M. A. McLaughlin, et al., ApJ, 473 (1996) 763. 
\bibitem{wall00}P. Wallace, et al., ApJ, 540 (2000) 184. 
\bibitem{matt97} J. R. Mattox, et al., ApJ, 476 (1997) 692. 
\bibitem{wehr97} A. Wehrle, et al., ApJ, 497 (1998) 178. 
\bibitem{nand97} G. Nandikotkur, et al., AIP Conf. Proc. 410 (1997) 1361.  
\bibitem{wagn97} S. J. Wagner, C. von Montigny, M. Herter, 
AIP Conf. Proc. 410 (1997) 1457. 
\bibitem{shra97} C. R. Shrader \& A. E. Wehrle, AIP Conf. Proc. 410 (1997) 328.
\bibitem{samb99} R. Sambruna, (1999), Proc. of the GeV - TeV Astrophysics 
International meeting, Snowbird, Utah; astro-ph/9912129.
\bibitem{foss99} G. Fossati, et al., MNRAS, 299 (1998) 433. 
\bibitem{ghis98} G. Ghisellini et al., MNRAS, 301 (1998) 451. 
\bibitem{urry99} C. M. Urry, APh, 11 (1999) 159. 
\bibitem{week99} T. C. Weekes, (2000), Proceedings of the International 
Symposium on High Energy Gamma-Ray Astronomy, Heidelbergl; astro-ph/0010431. 
\bibitem{kata99} J. Kataoka, et al., ApJ, 514 (1999) 138. 
\bibitem{sree99b} P. Sreekumar, et al., AIP Conf. Proc. 510 (1999) 318. 
\bibitem{liny92} Y. C. Lin, et al., ApJ, L401 (1992) 61. 
\bibitem{schl96} R. Schlickeiser, Space Sci. Rev., 75 (1996) 299. 
\bibitem{blan79} R. D. Blandford, \& A. K\"onigl, ApJ, 232 (1979) 34
\bibitem{boet99a} M. B\"ottcher, (1999) Proc. of the GeV - TeV Astrophysics 
International meeting, Snowbird, Utah; astro-ph/9909179. 
\bibitem{boet99} M. B\"ottcher, ApJ, L515 (1999) 21
\bibitem{hart00} R. C. Hartman, et al., (2000), ApJ, submitted. 
\bibitem{mann93} K. Mannheim, A\&A, 269 (1993) 67. 
\bibitem{mann96} K. Mannheim, Space Sci. Rev., 75 (1996) 331. 
\bibitem{ahar00} F. A. Aharonian, New Astronomy, 5 (2000) 377. 
\bibitem{muck00} A. M\"ucke \& R. Protheroe, APh, accepted
(2000); astro-ph/0004052. 
\bibitem{prot00} R, Protheroe \& A. M\"ucke, astro-ph/0011154. 
\bibitem{chia98} J. Chiang, \& R. Mukherjee, ApJ, 496 (1998) 752. 
\bibitem{sree98} P. Sreekumar, et al., ApJ, 494 (1998) 523. 
\bibitem{muck98} A. M\"ucke \& M. Pohl,  AIP Conf. Proc. 410 (1997) 1233. 
\bibitem{stec96} F. W. Stecker \& M. H. Salamon, ApJ, 464 (1996) 600.
\bibitem{kerr95} A. D. Kerrick, et al., ApJ, 452 (1995) 588. 
\bibitem{goul66} R. J. Gould \& G. Schreder, Phys. Rev. Lett., 16 (1966) 
252. 
\end{references}
\end{document}